\documentclass[iop]{emulateapj}

\usepackage{amsmath}
\usepackage{graphicx}
\usepackage{natbib}
\usepackage{rotating}
\usepackage{xspace}
\usepackage{epsfig}
\usepackage{multirow}

\newcommand{\Msun}{{\rm M}_{\odot}/{\rm h}}
\newcommand{\Mvir}{{\rm M_{vir}}}
\newcommand{\Rvir}{{\rm R_{vir}}}

\newcommand{\SUB}{{\small{SUBFIND}}\xspace}
\newcommand{\FOF}{{\small{FOF}}\xspace}
\newcommand{\FOFs}{{\small{FOF}}s\xspace}
\newcommand{\brems}{{bremsstrahlung}\xspace}
\newcommand{\Nmost}{{100}\xspace}
\newcommand{\prev}{{`prev'}\xspace}
\newcommand{\next}{{`next'}\xspace}
\newcommand{\dphi}{{\Delta E/E}}
\newcommand{\LCDM}{{$\Lambda$CDM}\xspace}

\begin{document}

\title{A First Look at Galaxy Flyby Interactions: I. Characterizing the Frequency of Flybys in a Cosmological Context}
\author{Manodeep Sinha\altaffilmark{1}, and Kelly Holley-Bockelmann\altaffilmark{2}}
\affiliation{Department of Physics and Astronomy, Vanderbilt University, Nashville, TN, 37235}
\altaffiltext{1}{manodeep.sinha@vanderbilt.edu}
\altaffiltext{2}{k.holley@vanderbilt.edu}

\begin{abstract}
Hierarchical structure formation theory is based on the notion that mergers drive galaxy evolution, 
so a considerable framework of semi-analytic models and N-body simulations has been constructed to 
calculate how mergers transform a growing galaxy. However, galaxy mergers are only one type of major 
dynamical interaction between halos -- another class of encounter, a close flyby, has been  largely ignored. 
We use cosmological N-body simulations to reconstruct the entire dynamical interaction history
of dark matter halos. We present a careful method of identifying and tracking a dark matter halo which 
resolves the typical classes of anomalies that occur in N-body data.  
This technique allows us to robustly follow halos and several hierarchical levels of subhalos as they grow, dissolve, 
merge, and flyby one another -- thereby constructing both a census of the dynamical interactions in a volume and an 
archive of the dynamical evolution of an individual halo. In addition to a census of mergers, our tool  characterizes 
the frequency of close flyby interactions in the Universe. We find that the number of close flyby interactions is 
comparable to, or even surpasses, the number of mergers for halo masses $\gtrsim 10^{11}\,\Msun$ at $z \lesssim 2$. Halo 
flybys occur so frequently to high mass halos that they are continually perturbed, unable to reach a dynamical equilibrium. In 
particular, we find that Milky Way type halos undergo a similar number of flybys as mergers irrespective of mass-ratio for $z\lesssim 2$.
We also find tentative evidence that at high redshift, $z \gtrsim 14$, flybys are as frequent as mergers. Our results 
suggest that close halo flybys can play an important role in the evolution of the earliest dark matter halos and their 
galaxies, and can still influence galaxy evolution at the present epoch. Our companion paper quantifies the 
effect of close flyby interactions on galaxies and their dark matter hosts.
\end{abstract}


\maketitle

\section{INTRODUCTION}
In a $\Lambda$CDM Universe, the smallest dark matter halos form first; bigger halos are then formed
via successive mergers with smaller halos. Thus, mergers are instrumental in the formation and evolution
of halos.  Galaxy mergers are rare, punctuated events in a galaxy lifespan that nonetheless dramatically 
change it -- from its morphology~\citep[e.g.,][]{H41,TT72,S86,BH92, BH96, MH96,MKLD96,SW99, DMH99, B02,
SDH05,C06,HH06,RBCD06,R06,NJB06,NKB06,CD08,MMW98,HR00,B02}, to its stellar population~\citep[e.g.,][]{MH94,HM95,CJ08,BK08}, 
to the evolution of the central supermassive black hole~\citep[e.g.,][]{SMH05,HH05,HH06,MHSA07,YHCH08, MHS11}. 
Consequently, merger rates have been studied extensively, both  theoretically~\citep[e.g.,][]{LC93,GW08,GGBS08,GGBN09} 
and observationally~\citep[e.g.,][]{S86, R02, L03, CBG04,VD05,BACC07,LPKC08,TGSN08,ULB08,RCWS08}. Collisionless cosmological N-body
simulations can be used to measure halo merger rates, where a merger is defined to occur when a bound dark matter halo
falls into another bound dark matter halo. Various simulations measuring such halo merger rates agree
to within a factor of $\sim 2$~\citep[see][for a discussion on merger rates from collisionless simulations]{HC10}. Galaxy merger
rates can then be inferred from the subhalo mergers within a primary halo by assuming a $M_{\rm halo} - M_{\rm gal}$ 
relation~\citep[e.g.,][]{GW08,WCW09,BCW10} or directly measured in hydrodynamic simulations~\citep[e.g.,][]{MKKDW06,SWDGKK09}. 
Observationally, merger rates are typically derived from close-pair counts -- i.e. galaxies with small projected separations 
and relative velocities -- and are globalized using an estimate of the lifetime or duration of the observed merger phase~\citep{LJCP10}. 

Ultimately, galaxy mergers are successful in shaping galaxy properties
because they cause a large perturbation within the potential. Even an
orbiting satellite can distort the underlying smooth galaxy potential
and produce observable effects. For example, the HI warp in the Milky
Way disk may be tidally triggered by the Large Magellanic
Cloud~\citep{WB06}. However, one entire class of galaxy interactions
also capable of causing such perturbations -- galaxy flybys -- has
been largely ignored.

Unlike galaxy mergers where two galaxies combine into one remnant, flybys occur when two independent galaxy halos interpenetrate 
but detach at a later time; this can generate a rapid and large perturbation in each galaxy. We developed and tested a method to 
identify mergers and flybys between dark matter halos in cosmological simulations and to construct a full `interaction network' 
that assesses the past interaction history of any given halo. With this new tool, we are undertaking the first systematic
study to quantify the frequency of flybys and its effect on galaxy evolution.

In this paper, we present our technique for determining the network of
dynamical interactions for halos in an N-body simulation, and present
a census of halo flybys and mergers in the Universe.  We discuss our
simulation and halo-finding technique in Section 2.  Section 3 describes
our technique to identify halo flyby interactions, and to construct a
halo interaction network. Section 4 presents the results, and Section
5 discusses the implications of flyby interactions and previews the
next paper in the series. The appendix covers our method of
linking parent and child halos, including ways to
mitigate common problems that plague this process.

\section{SIMULATION TESTBED AND HALO IDENTIFICATION}\label{sec:halo_identify}
We use a high-resolution, dark matter simulation with $1024^3$ particles in a box of length 50 Mpc/h 
with {\small WMAP-5} cosmological~\citep{K08} parameters as a testbed to develop our technique.  The
initial particle distribution is obtained from a Zeldovich linear approximation at a starting redshift of $z=249$ and is evolved using the 
adaptive tree-code, {\small GADGET-2}~\citep{SYW01,S05}. The dark matter particles have a fixed gravitational co-moving softening length of 
$2.5$ kpc/h. We store 105 snapshots spaced logarithmically in scale-factor, $a = 1/(1+z)$, from $z=20$ to $0$. This translates
into a timing resolution $\lesssim 50$ Myr for $z\gtrsim 3$ and $\sim 150$ Myr for $z \lesssim 3$. 
Since the fundamental mode goes non-linear at $z=0$, we will only present results up to $z=1$ where the 50 Mpc/h
box is still a representative cosmological volume. 

In addition, we ran two simulations designed to explore the effect of mass 
and timing resolution on the flyby phenomenon. These two volumes, with $512^3$ 
and $1024^3$ particles respectively, have the same size and cosmology as our testbed 
simulation and were evolved from $z=249-1$. The particles were drawn from 
identical phases, and with 161 snapshots total, had timing resolution better than 200 Myrs. 

To begin identifying halos, we first use a Friends-of-Friends(\FOF) technique with a 
canonical linking length $b=0.2$ ($\sim 10$ kpc/h)~\citep{DEFW85}. 
We require at least 20 particles  ($\sim 10^8 \Msun$) to define a halo, but our halo 
interaction network uses only those halos with greater than 100 particles; we discuss the effects of 
this limit in Section~\ref{section:resolution}. Subhalos (down to multiple hierarchy levels) are 
identified using the \SUB algorithm \citep{SWTK01}. The \SUB algorithm identifies subhalos as bound 
structures around a density maxima\footnotemark.
The remaining particles in the \FOF halo are comprised of particles that are bound to the potential of the main halo but not bound to any
subhalo. For the remainder of this paper, all references to the {\it main} or {\it primary} halo will mean this bound set 
of ``background" particles. For details on the \SUB algorithm, we refer the reader to \citet{SWTK01}. We caution that \SUB, 
or any density-based halo extraction technique, only recovers subhalo masses enclosed within a region
of higher subhalo density compared to the background~\citep[see][]{MPP10}. Therefore, a subhalo will lose mass depending on the 
radial distance from the center of the main halo; i.e., the subhalo mass is only to be trusted modulo a density contrast. If we 
want a detailed look at the subhalo mass loss evolution, then we must carefully correct for this bias. Fortunately, at this 
stage we simply require the existence of the subhalo; we caution that there is one consequence of this bias that presents a 
challenge: in the most extreme case, a subhalo passing through the center of the main halo can disappear for multiple 
snapshots only to reappear later, where it can be mistaken for a new subhalo. We account for such missing subhalos while 
constructing the halo interaction network (see Section~\ref{sec:network}). Since our focus is on flybys, we relegate most 
of the discussion of our technique to the Appendix, including resolution tests and comparison to other techniques.

\footnotetext{A particle is associated with only one subhalo}

\section{TAXONOMY OF HALO DYNAMICS}
In a collisionless simulation, dark matter halos can grow via mergers or through smooth accretion. They can also lose mass through dynamical 
interactions, such as flybys or 3-body encounters~\citep[e.g.][]{SNAS07,LNSJFH09}; primary halos can even be stripped of entire subhalos.
Halos close to the numerical resolution limit can disappear for multiple snapshots. A primary halo can  fall into another primary halo 
and continue to survive as a subhalo that is subsequently stripped. Ideally, a robust account of a halo's dynamical past captures all 
such scenarios and produces a physically meaningful history for individual halos. In a nutshell, our halo interaction network 
attempts to capture all of these interactions. We construct our halo interaction network in a three-step fashion: 1) for every halo, 
we find the best child halo at some later snapshot (if it exists); 2) we trace the complete set of parent-children pairs through 
all snapshots to construct the halo network; and 3) using the phase space history, we characterize the type of interaction for each pair. 

Capturing the rich history of interactions of dark matter halos requires identifying every 
possible interaction. Most of the interesting phenomena begin with a main halo falling into another main
halo. Assuming that a secondary main halo falls into a primary main halo, the following outcomes are 
possible:
\begin{itemize}
\item The secondary (now a subhalo) continues to orbit inside the  main halo and ultimately
disrupts inside that same main halo (merger started). 
\item The secondary continues to orbit the primary halo and exists at $z=0$ (merger completed).
\item The secondary merges with another subhalo inside the same primary halo (subhalo merger).
\item The primary main becomes a subhalo itself by merging with another halo. The secondary is now a sub-subhalo (merger).
\item The secondary halo enters and then detaches from the primary halo (grazing flyby).
\item Either the secondary or the primary disappear (transients)
\end{itemize}
It is also possible that a subhalo appears without ever having been a main halo. Based on our
snapshot timing resolution work (see Appendix), we find that these subhalos formed 
as main halos very close to the primary and merge in between snapshots. 

An interaction begins the moment a halo changes its state, e.g., when a main halo becomes a subhalo of 
another main halo. We categorize each interacting pair of halos 
by tracking the behavior of the pair into the future. At the moment the encounter begins, we
tag the interaction according to the encounter type. For instance, when a subhalo enters a primary 
and eventually disrupts inside it, the interaction is tagged as a merger. Along with this tag, we store
the redshift of infall (when the subhalo first becomes part of the main) and the destruction redshift. 
Thus, we automatically get a duration for every interaction. In addition, we store the minimum separation 
between the subhalo and primary halo and the normalized radius at destruction -- this allows us to take 
census of the galactocentric distances reached by any particular interaction as a function of primary mass, 
mass ratio and infall redshift while accounting for the dynamic behavior of the primary halo. 

Since we use the redshift of infall as interaction redshift, mergers and flybys in our simulation will occur well
before any observable encounter-induced features within the galaxies. For instance, the peak in the merger rate occurs 
for $z \sim 5$ (see Fig.~\ref{fig:flyby_freq}) instead of $z \approx 2$ as found elsewhere~\citep{MFD96,RCWS08}. Note, though, 
that this definition of recording a merger as the redshift of infall is in line with the techniques used in the literature to 
construct mergertrees using \FOF halos. 

\subsection{Examples of halo evolution within the network}
\begin{figure*}[ht]
\begin{minipage}[t]{0.7\linewidth}
\centering
\includegraphics[height=7.6in,keepaspectratio=true,clip=true]{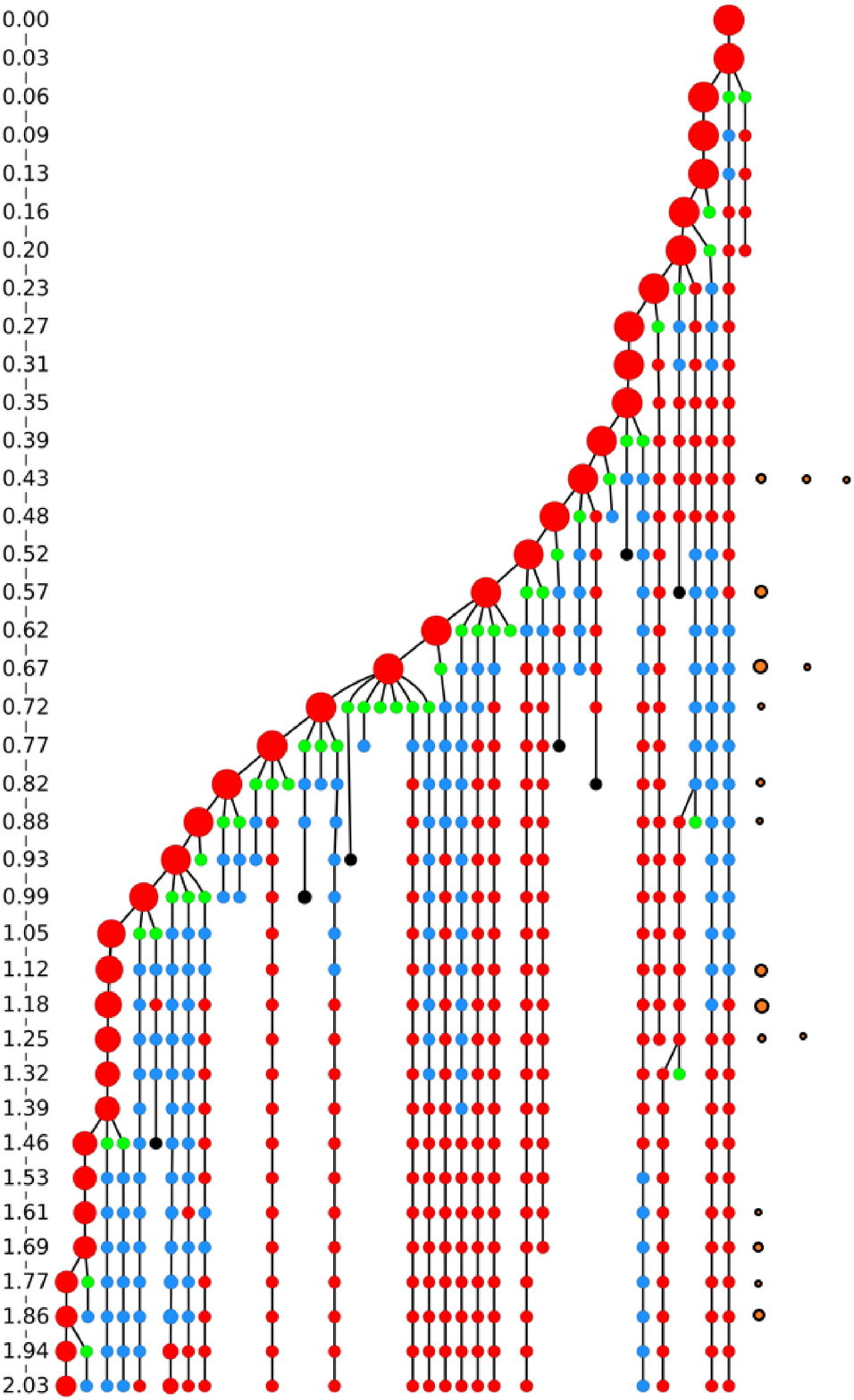}
\end{minipage}
\hspace{0.2cm}
\begin{minipage}[t]{0.3\linewidth}
\centering
\includegraphics[height=7.6in,keepaspectratio=true,clip=true]{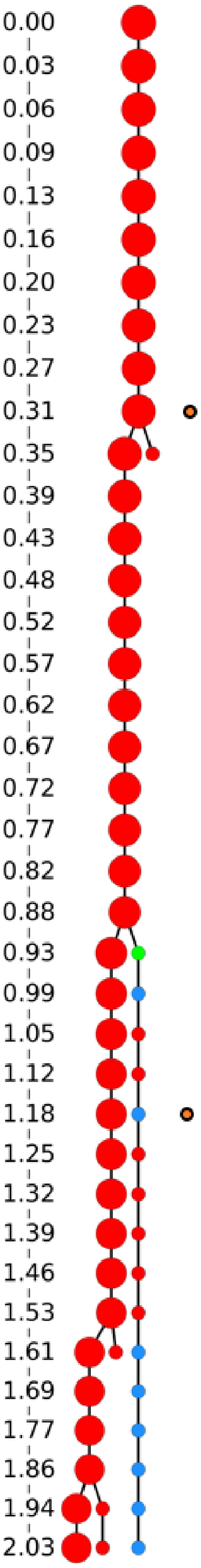}
\end{minipage}
\vspace{0.1cm}
\caption{\small {\em Left}: Merger tree for a hectic halo.  The mass of the halo at $z=0$
and $z=4.5$ is $1.8\times10^{12}\,\Msun$ and $2\times10^{10}\,\Msun$ respectively.
Red represents a primary halo, blue represents a subhalo, 
green represents a subhalo that is going to dissolve inside the main halo, and black represents a 
halo that skips at least one snapshot. The size of the circle
represents the halo's log-scaled mass with respect to the mass of the final main halo at $z=0$. We only show those halos 
that are at least  $1/10$ of the primary mass. The corresponding redshifts are shown on the left. 
The flybys are shown on the right with yellow circles; the size of the circle is scaled to show
the mass ratio of the flyby. The main halo undergoes 77 mergers and 21 flybys; however, we can only see a specific type of flyby in this 
mergertree -- those relatively rare flybys that eventually result in a merger. 
{\em Right}: Merger tree for a quiet MW mass halo. Flybys are shown as yellow circles on the right. 
If a subhalo actually survives until $z=0$, then it will not appear in this mergertree. At $z=0$, the halo mass is 
is $10^{12}\,\Msun$ , while at $z=4.5$, it is $2\times10^{11}\,\Msun$; this halo assembled most of its mass early via smooth accretion.}
\label{fig:dot_images}
\end{figure*}

In Fig.~\ref{fig:dot_images} we show the halo interaction network for two Milky-Way type 
halos of $\sim 10^{12} \,\Msun$ at redshift zero\footnotemark. These are two drastically different halos, even 
though they have the same mass at $z=0$. The first halo formed at $z=9.3$ and had a very active merger history; 
by $z=0$, the assembly of this halo required 434 mergers. The second halo formed earlier, at $z=12.5$, yet only 
required 111 mergers -- a `quiet' evolution. Much of the mass in the quiet halo is assembled early through 
smooth accretion of dark matter; at $z\sim 5$, the hectic halo has a mass of $2\times10^{10}\,\Msun$ while the
quiet halo is 10 times more massive.   The hectic primary halo itself directly survives 77 mergers, while 
the quiet halo undergoes only 27. Such divergent pasts can leave an imprint on the structure
and properties of the galaxies contained in these halos -- in the star formation history, the 
morphology, and perhaps the central black hole mass. We will explore this topic in a separate 
paper (Sinha \& Holley-Bockelmann, in prep). 

\footnotetext{We used the total mass of the main halo, including all the subhalos, as proxy for the virial mass, 
and assumed a spherical halo to get the virial radius. }

\subsection{Finding Flybys}
One major goal of this paper is to characterize flyby interactions -- interactions that do not end with one halo 
accreting another. In principle, there are three classes of close flyby interactions:
\begin{itemize}
\item {\bf Grazing} -- Two primary halos approach on convergent trajectories, interpenetrate
for at least half a crossing time, and then separate as two distinct primary halos once more. 
\item {\bf External} --  Same as above but the primary halos remain distinct at all times. This can also apply to halos that are at the 
same hierarchy levels within the same container halo (e.g., two subhalos of the same main halo). Since 
technically this can include every other halo in the volume, it is useful to define a maximum pericenter distance between two halos when classifying this type.
\item {\bf Internal} --  A halo at a higher hierarchy level passes close to the center of its containing halo, e.g., 
a sub-sub halo goes through the central regions of its containing subhalo. This is synonymous with
the decay of satellite orbits after a merger.
\end{itemize}
Internal flybys have been studied in great detail~\citep[e.g.,][]{TO92,QHF93,WMH96,SNT98,J98,TB01,PKB09,KZKBD09}, so we will not 
discuss them here. External flybys are naturally distant encounters, where the separation between halo centers, ${\rm R_{sep}}$, is 
larger than the virial radius of each halo. Since the perturbation in the potential induced by a halo flyby 
is $\propto {\rm R^6_{sep}}$~\citep{VW00}, external flybys excite comparatively weak perturbations in the galactic 
potential. Since we are interested in potentially transformative interactions, we 
will focus on grazing flybys. For the remainder of this paper, a `flyby' means a grazing encounter.

\begin{figure}
\centering
\begin{tabular}{cc}
\includegraphics[width=0.48\linewidth,clip=true,keepaspectratio=true]{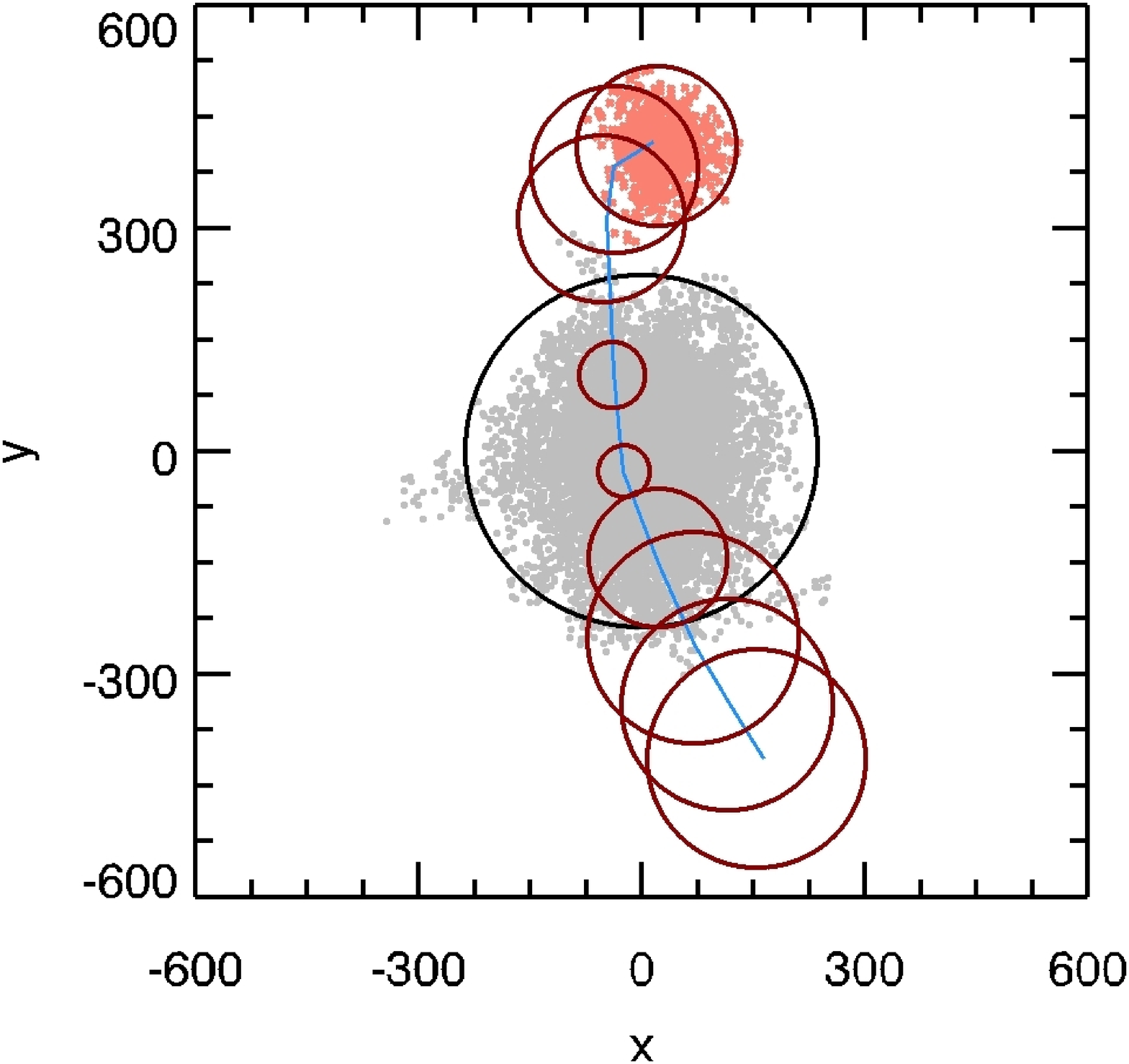}&
\includegraphics[width=0.48\linewidth,clip=true,keepaspectratio=true]{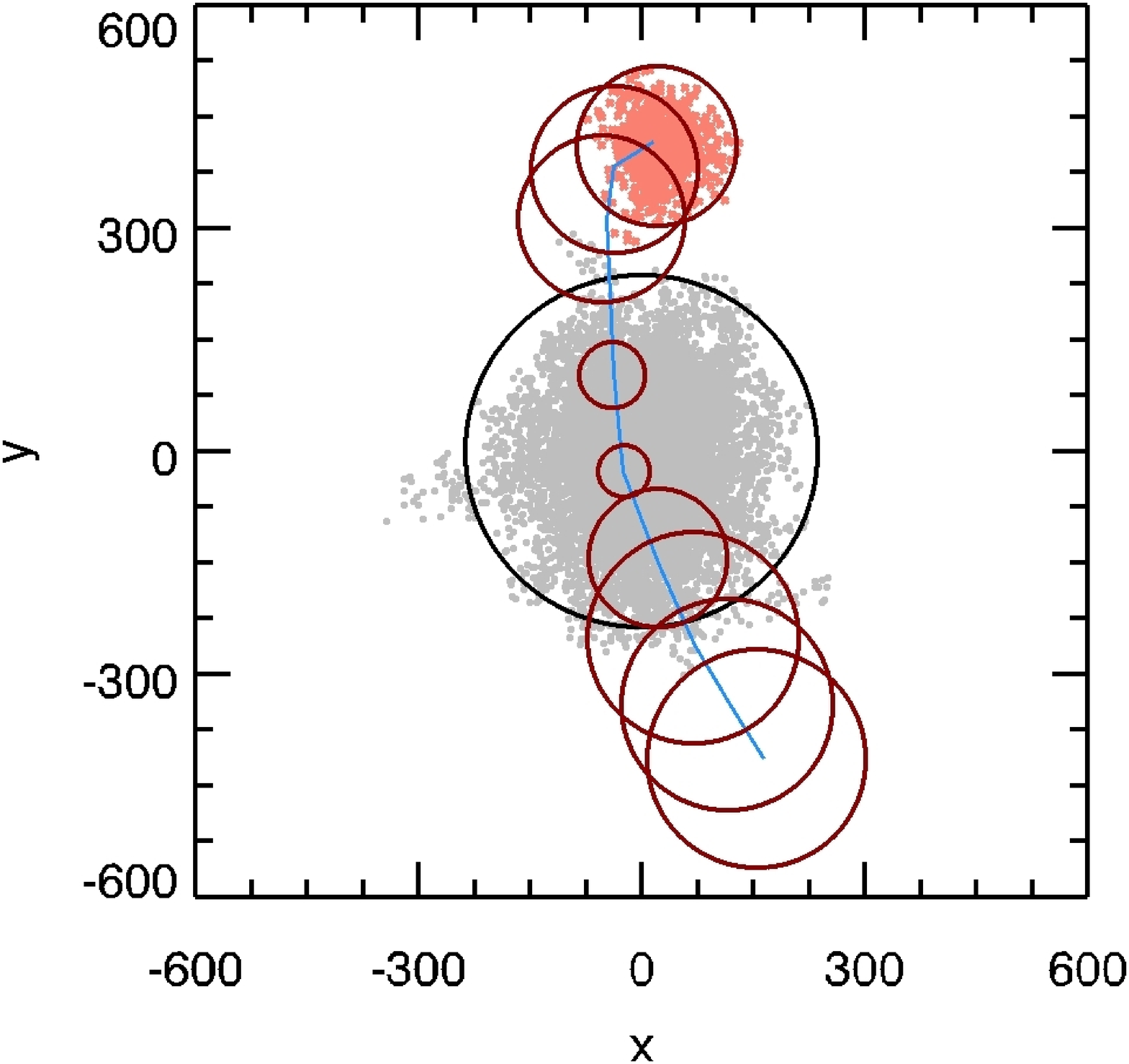}\\
\includegraphics[width=0.48\linewidth,clip=true,keepaspectratio=true]{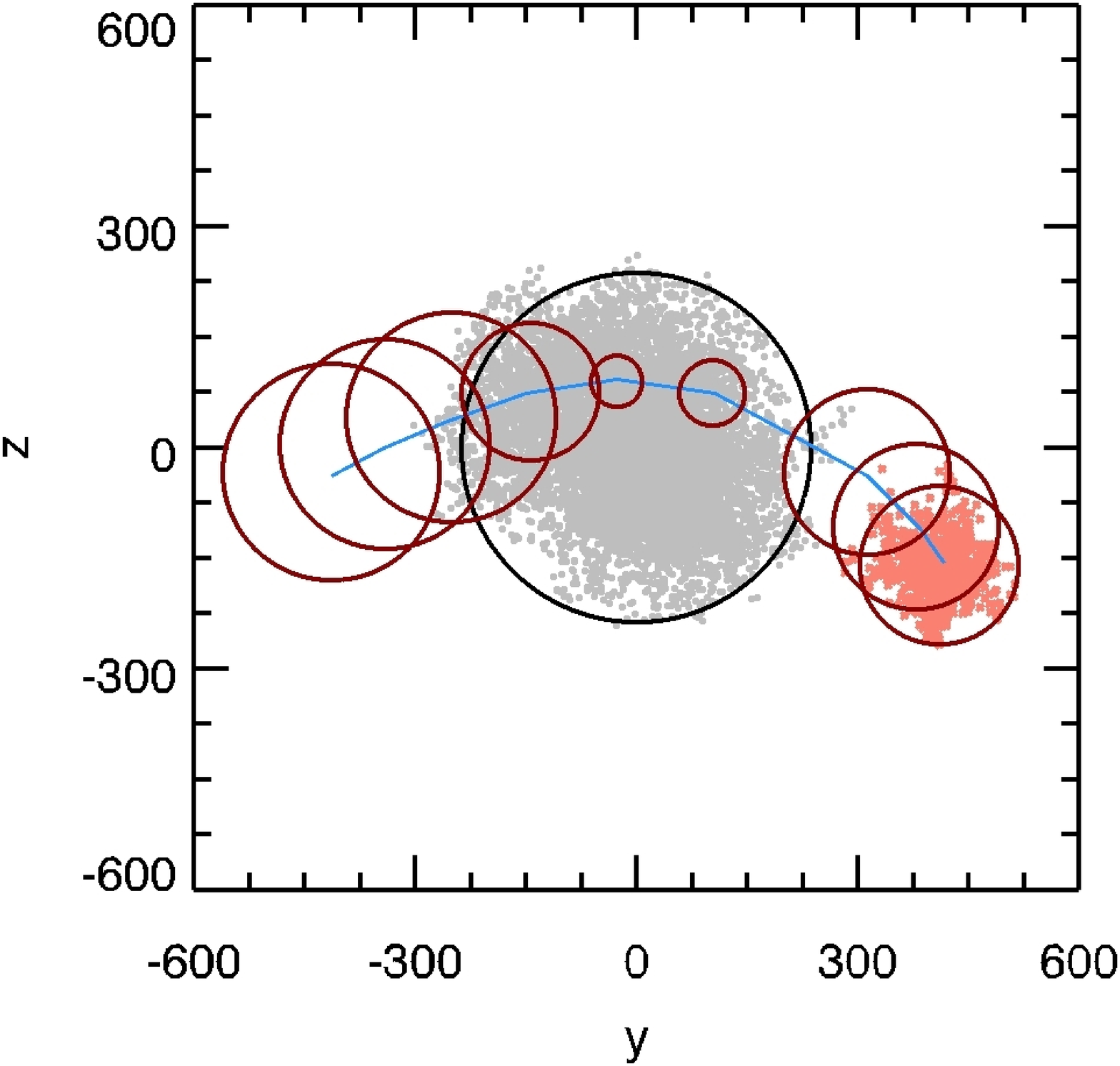}& \\
\end{tabular}
\caption{\small This figure shows a 5:1 flyby between a Milky Way type halo and a satellite halo. The flyby starts 
at $z=1.1$ and lasts until $z=0.3$, for a duration of $4.7$ Gyr. The gray and red points show the particle distribution 
of the \FOF and the subhalo. The black and the dark red circles show the virial radius of the main halo and 
the subhalo throughout the encounter while the blue line shows the trajectory of the subhalo center of mass. This figure also 
shows the artificial mass-loss experienced by subhalos as they pass close to the main halo center. }
\label{fig:mw_flyby}
\end{figure}

We require that the subhalo in flyby remain in the primary halo for at least half a crossing time at
the time of infall. The crossing time, $t_{\rm cross} = \sqrt{{\Rvir^3}/{\rm G \Mvir}}$, is independent of mass. Because 
we define halos by an overdensity, $\Delta_{\rm vir}  = 200 \times \rho_{\rm crit}$, $t_{\rm cross}$  simplifies 
to $0.1 \times H(z)^{-1}$. Our flyby definition is physically motivated in that we concentrate on longer-duration flybys 
that are more likely to leave an lasting imprint on the structure of the halos. 
However, this definition does exclude some rapid, transient events -- hence, the results presented in this paper are 
conservative estimates of the true flyby rate in the Universe. 

For grazing encounters,  the interaction will always be a flyby for very large relative velocities ($v_{\rm rel} >> v_{\rm esc}$ of the 
combined halo masses). However, we further compute the velocity dot product for the main halo
centers in the center of mass frame at the snapshot where the halos are separate, $ (v_1 - v_\mu)\cdot (v_2 - v_\mu)$. 
If the dot product is negative, then the halos were approaching before they interpenetrate. We found that
the flybys with negative $\mathbf{v\cdot v}$ are slower and sink deeper into the main halo than the flybys with positive $\mathbf{v\cdot v}$. 
Fig.~\ref{fig:mw_flyby} shows an example of a deep flyby with a mass ratio of 5:1 from the testbed simulation. The event begins
at $z=1.1$ and completes by $z=0.3$ --  a duration of $\sim 4.7$ Gyr. At infall, the main halo and subhalo virial radii are 
$\sim$ 260 and 150 kpc/h respectively. We find the two halo centers are closest at $z \sim 0.8$ with a separation 
of $\sim 100$ kpc/h; the subhalo itself overlaps the main halo center here, which suggests that this encounter can perturb even
the innermost regions of the main halo. Our technique is designed to identify this type of strongly-interacting, long-duration 
flyby, as it has the most potential to transform the overall structure of each halo. 

\section{RESULTS}
\subsection{Mass and time resolution effects}
In this section, we estimate the effect of particle resolution and snapshot timing 
resolution on the flyby rate. We have used a 512$^3$ simulation and a 1024$^3$ simulation 
that is exactly the same phase for the results in this section. 

\begin{figure}
\centering
\includegraphics[width=0.48\textwidth,keepaspectratio=true,clip=true]{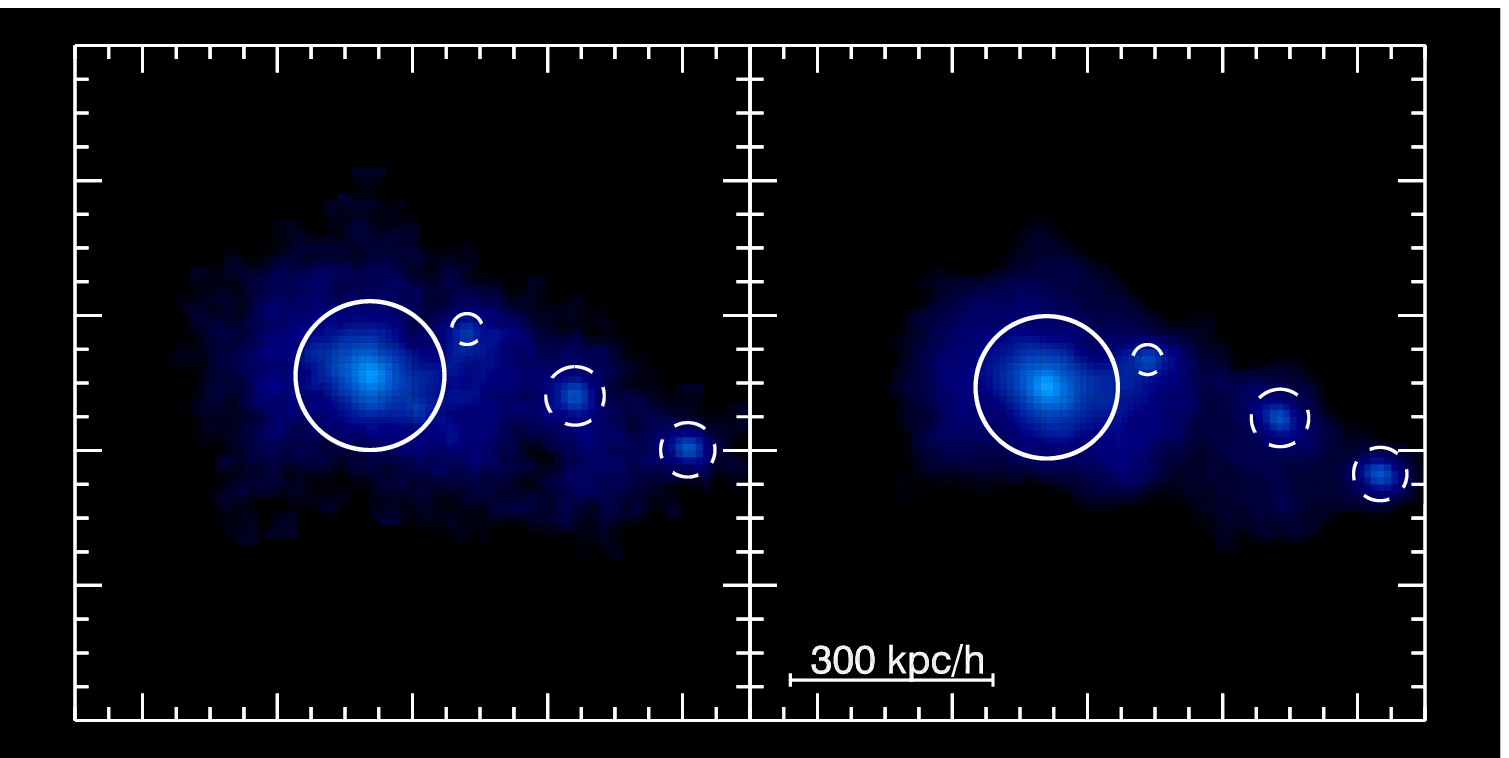}
\caption{\small The solid white circle marks the virial radius of a typical halo extracted from the $512^3$ simulation ({\it left}) and the 
$1024^3$ simulation ({\it right}) at $z\sim 1.8$. Dashed white circles locate the other \FOFs in the volume. Identical cubes of side 1 Mpc/h were extracted from both the simulations
and projected to make this plot. This halo experienced the same number of merger and flybys throughout the run.}
\label{fig:mw_diffphases_image}
\end{figure}

In order to determine the sensitivity of flyby rates to numerical resolution, we 
conducted two experiments that degraded our fiducial simulation. In the first experiment, 
we explored the effect of particle number on the frequency of flybys with a $512^3$ simulation 
that was seeded with precisely the same initial random phases as the $1024^3$ run~(Rom\'an Scoccimarro, private communication). 
Figure~\ref{fig:mw_diffphases_image}  centers on a typical MW halo in the volume -- 
this halo experienced no difference in the number of mergers or flybys over its entire history.
In the entire volume, the difference between the number of flybys in the two simulations is less than $1$ percent, 
when we consider the halos that are well-resolved in both simulations (i.e., down to the minimum halo mass in the 
512$^3$ volume of  $M  < 6.7\times 10^{10} \Msun$). 
In addition, the number of flybys in two-dimensional mass and redshift bins are identical within Poisson error for $95\%$ of the
bins for these two simulations. 
So we find that particle resolution does not affect flyby rates. 

The second series of tests concerned the timing resolution of the simulation. In the 512$^3$
simulation, we wrote snapshots in equal increments of log $a$, but never allowed more than 200 Myr between 
snapshots -- this upper limit ensures that encounters for MW mass objects were resolved over the 
course of the run. With 161 snapshots, we made two new halo interaction network by skipping one and 
two snapshots respectively. Thus, we have three samples built from 161, 80 and 54 snapshots
respectively to quantify the effects of snapshot timing resolution on the flyby rate. 
We analyze these three samples and compare the total number of flybys between halos. 
We find that to get convergence in the flyby rate, the snapshot outputs need to resolve the `typical'
flyby duration. Since a flyby is tagged as a main halo $\rightarrow$ subhalo $\rightarrow$ main halo transition, and we 
require flybys to last at least half a crossing time, the snapshots must be able to resolve events on 
timescales of $\sim t_{\rm crossing}$. For the simulations with 161 and 80 snapshots, the crossing time 
was always resolved by snapshots and resulting number of flybys differ by less than 1\% and are within
Poisson errors. However, the 54 snapshot simulation did not resolve the crossing time for most 
redshifts, and as a result the number of flybys is $\sim 20\%$ lower. 

As another test, we re-simulated the volume between $z=3-2$ with a snapshot timing resolution of 25 Myrs, which yielded 
45 snapshots. The crossing times at $z=3$ is $\sim$ 330 Myrs, so it is very well resolved in this simulations. 
We take a similar approach as above, creating a series of progressively coarser halo interaction networks by skipping 
over snapshots. As before, we find that the 
number of flybys is consistent within Poisson error when we resolve the crossing time 
and drops dramatically for coarser simulations. Thus, our definition of flybys produces consistent flyby rates as 
long as the snapshot outputs can resolve the crossing time. Otherwise, the flyby rates reduce as the time resolution 
degrades. This implies simulations with insufficient outputs will underestimate 
the true flyby rate. Assuming that the time between snapshots is at most a quarter of the crossing 
time (such that flybys last for at least 2 snapshots), we find that the total number of snapshots 
required between redshifts 20 and 0 to be $\sim$ 132. This is in agreement with recent findings that show 
at least 128 snapshots are required to obtain a robust estimate of halo masses~\citep{BBDBM11}. 

\subsection{Frequency of flybys}
Now that we have a way to track and classify interactions between halos, it is useful to determine
how common close flyby encounters are compared to mergers. This is an important first step 
in determining whether current semi-analytic and high resolution N-body models could be 
missing a significant number of galaxy-transforming interactions. Note that in this census, we are simply 
keeping a tally of dark matter halo mergers and close halo flybys, regardless of how much they may perturb 
the potential of any galaxy embedded within. In the next paper, we will concentrate on quantifying the perturbation 
caused by these types of encounters. We checked to see if there are multiple flybys between the same halo pairs
and found that $\sim$ 70\% of the flybys do not recur. About 20\% of flybys eventually become a merger,  $\sim$ 6\%
are repeat flybys. Thus, flybys primarily represent one-off events between two halos. 

\begin{figure}
\centering
\includegraphics[scale=0.12,clip=true]{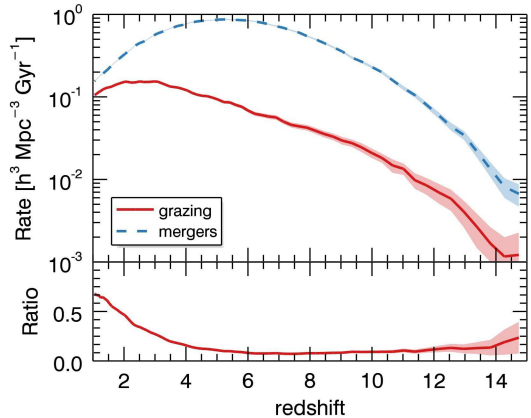}
\caption{\small The top panel shows the number of mergers (blue, dashed line) and flybys (red, solid line) per co-moving Mpc$^3$ per Gyr as a
function of redshift. The shaded region shows the Poisson error on the number of interactions. 
The bottom panel shows the ratio between the number of flybys to mergers vs redshift. 
Mergers dominate flybys in the redshift range $\sim 4-12$ but flybys become relatively more common 
at $z\lesssim 3$. We can also see tentative evidence for an increase in the relative fraction of flybys
compared to mergers at $z \gtrsim 14$. However, this increase is also due to the low number of interactions 
and is borne out by the large Poisson error at those high redshifts. }
\label{fig:flyby_freq}
\end{figure}

In Fig.~\ref{fig:flyby_freq}, we present the number of flybys and
mergers per Gyr per cubic Mpc as a function of redshift.  Mergers dominate
the overall rate of interactions for $12 \lesssim z \lesssim 4$ by about an order
of magnitude. However, at $z\sim 14$, we see tentative evidence of an increase
in the relative number of flybys compared to mergers. At low redshift, $z \lesssim 3$, 
flybys start to become more prevalent. The impact of this on observed merger rate estimates may be profound: in large-scale surveys, the rate of 
close-pair galaxies is often taken as a proxy for merger rate, but if we assume that 
50\% of these halo flybys appear as {\em galaxy} flybys, then Fig.~\ref{fig:flyby_freq}
implies that at $z\lesssim 3$, such galaxy-pair counts are going to be contaminated by 
flyby-halos at least at a 20-30\% level. Flybys also result in bimodal kinematic 
galaxy distribution near the virial radius, in that some galaxies are infalling for the
first time, while other galaxies that were once much deeper inside the cluster potential 
are now leaving. Conceptually, these flyby galaxies are similar to `backsplash galaxies'~\citep[e.g.,][]{GKG05,MDS04,KLKMYGH11}. 

\begin{figure}
\centering
\includegraphics[width=0.45\textwidth,keepaspectratio=true,clip=true]{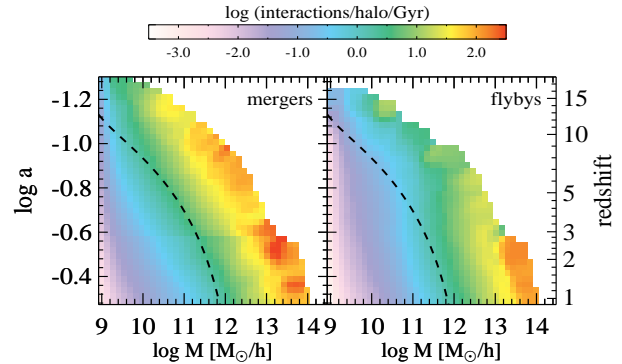}
\caption{\small This figure shows the number of mergers (left) and flybys (right) per halo per Gyr as a function of primary halo mass and redshift. 
The dashed line shows the mass accretion history of a typical Milky-Way type obtained from our simulations~\citep{WBPKD02}. The number
of flybys increases with the primary halo mass for all redshifts, consistent with \LCDM. For $z\lesssim 3$, halos above $10^{12}\,\Msun$ 
are have flyby rates greater than 100 per Gyr. Such a high flyby rate (in addition to mergers) means these halos are unlikely to be in
equilibrium. }
\label{fig:rate_interactions_gyr_primary}
\end{figure}

\begin{figure}
\centering
\includegraphics[width=0.45\textwidth,keepaspectratio=true,clip=true]{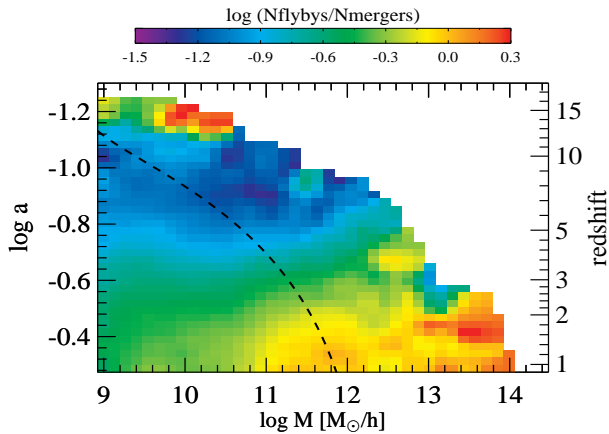}
\caption{\small Ratio of number of flybys to mergers as a function of primary halo mass and redshift. As we saw in Fig.~\ref{fig:flyby_freq}, 
mergers dominate flybys by an order of magnitude for $12\lesssim z\lesssim4$. At lower redshifts, however, flybys start becoming
more prevalent and by $z\sim 2$, flybys are at comparable or even larger for all halos above $10^{11} \,\Msun$.  }
\label{fig:flybys_over_nmergers_primary}
\end{figure}

While Fig.~\ref{fig:flyby_freq} shows the global flyby rate, we need
to know the rate of interactions as a function of both halo mass and redshift to assess the impact flybys
may have on halo evolution. Interaction rates can be expressed either as `per object' or `per volume'. Here, we plot the 
rates per halo to reduce the additional dependence of the halo mass function evolution. In Fig.~\ref{fig:rate_interactions_gyr_primary} 
we show the number of flybys and mergers on a per halo per Gyr basis, while the ratio of flybys to mergers is 
seen in Fig.~\ref{fig:flybys_over_nmergers_primary}.  We see that flybys are more frequent for higher-mass halos 
at all redshifts, consistent with the \LCDM  framework. From Fig.~\ref{fig:flybys_over_nmergers_primary}, we
can directly see that the number of flybys becomes comparable or larger than the number of mergers for halos $> 10^{11}\,\Msun$ 
for $z \lesssim 2$.  Such halos are expected to host galaxies -- and the effect of flybys should leave an imprint on
the observable properties.  We also find tentative evidence that flybys are comparable in number to mergers at the 
highest redshifts, $z\gtrsim 14$ -- however, the numbers are affected by Poisson error and we can not 
conclusively say that flybys dominate mergers at those highest redshifts. We are simulating a bigger boxsize 
to look at the relative importance of flybys for $z\gtrsim14$. 

\begin{figure}
\centering
\includegraphics[width=0.45\textwidth,keepaspectratio=true,clip=true]{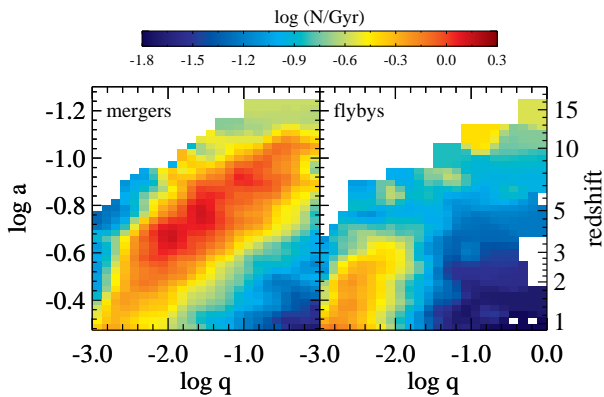}
\caption{\small The average number of interactions per Gyr for a MW type halo (selected at $z=0$) as a function
of mass-ratio ($q$) and redshift. Mergers are plotted on the left panel while flybys are on the right panel. As seen in the previous
plots, mergers dominate flybys for all $q$ values in the redshift range $12-2$ but at flybys become comparable in number at lower redshifts.}
\label{fig:mw_interactions_pergyr}
\end{figure}

In the preceding mergers versus flybys comparison figures, we used all interactions irrespective of mass-ratio, $q$. Now, it is 
interesting to see how the frequency of interactions compare as a function of both $z$ and $q$ for a typical MW halo. We select MW halos
at $z=0$ using three criteria: a) halo mass between 1 and 2 $\times 10^{12}\,\Msun$ and b) the halo has always been a primary halo 
and c) the main progenitor of the halo can be traced back to $z>6$. 
We found 178 such halos in our simulation box. In Fig.~\ref{fig:mw_interactions_pergyr}, we show the average frequency of interactions per Gyr 
for these halos as a function of $q$ and $z$. Mergers are dominant for $12 \lesssim z \lesssim 2$ for all $q$; however, flybys take over for 
lower $z$. For these MW type halos, {\em flybys and mergers happen with a similar frequency for $z\lesssim 2$} -- a trend that we 
saw in Fig.~\ref{fig:flybys_over_nmergers_primary}.

\subsection{Fits to interaction rates}
Merger rates are usually calculated in two different ways: as a rate i) {\em per parent halo} ii) {\em per child halo}. These two rates
physically mean slightly different things -- the first one is tied to the probability of a halo of a given mass at some $z$ having a merger. 
Observationally this is measured by the merger fraction through pair counts. The second one is the probability that a halo had a merger
in the past -- observationally identified by morphological features. Since we track subhalos, we can distinguish between the beginning of a merger
(i.e., a main halo$\rightarrow$subhalo transition -- tied to the merger rate per parent halo) and the end of a merger (subhalo$\rightarrow$main halo --
tied to the merger rate per child halo). These two methods of measuring the merger rate can yield somewhat different results~\citep[see][]{HC10}.
In this section we will use the time of primary$\rightarrow$subhalo transition as a merger time for the primary halo. 

To compare to merger rates found in the literature, we use
the analytic fit of~\citet{WCW09} and classify interactions per halo per Gyr in the form:
\begin{equation}\label{eqn:fits}
\dfrac{n_{\rm events}}{n_{\rm halo}{\rm dt}} = A (1+z)^\alpha,
\end{equation}
where, $n_{\rm events}$ is the number of mergers or flybys for a parent halo in some mass bin, $n_{\rm halo}$ is the number of halos in the same mass range
and ${\rm dt}$ is the time interval (in Gyrs) between consecutive snapshots. We consider three different mass ranges for the parent halos, $10^{10}-10^{11}$, 
$10^{11}-10^{12}$ and $10^{12}-10^{13}\,\Msun$. We limit the redshift range from $z=6$ to $1.0$. 
The fits for mergers and flybys are presented in 
Table~\ref{table:fits}. Overall, the values of $A$ for flybys and mergers are comparable, while 
$\alpha$ is higher for mergers. Correspondingly, the flyby rate increases more slowly than the merger rate with increasing $z$.

\scriptsize
\begin{table}[ht]
\caption{Fits to Eqn.~\ref{eqn:fits} for three main halo mass ranges. The redshift is noted when the subhalo first 
appears inside the main halo. The reduced $\chi$-squared values for the fits are also shown in the table. }
\centering
\renewcommand\arraystretch{1.6}
\begin{tabular}{cc|c|c|c}
\hline\hline
 &  ${\rm {\mathbf{M}}}_\odot$  & $\mathbf{10^{10}}$ -- $\mathbf{10^{11}}$ & $\mathbf{10^{11}}$ -- $\mathbf{10^{12}}$ & $\mathbf{10^{12}}$ -- $\mathbf{10^{13}}$ \\[0.5ex]\hline
\multirow{3}{*}{\textbf{Mergers}}
& A         & $1.33\times10^{-3}$    & $1.61\times10^{-2}$     &  $3.84\times10^{-1}$ \\
& $\alpha $ & 2.40                   & 2.58                    &  1.16 \\
& $\chi^2$  & 10.6                   & 3.05                    &  1.45  \\[0.7ex] \hline
\multirow{3}{*}{\textbf{Flybys}}
& A         & $1.08\times10^{-3}$    & $1.92\times10^{-2}$     & $2.59\times10^{-1}$ \\
& $\alpha $ & 1.02                   & 0.75                    & 0.68 \\
& $\chi^2$  & 2.33                   & 1.63                    & 0.88  \\[0.7ex]\hline\hline       
\end{tabular}
\label{table:fits}
\renewcommand\arraystretch{1.0}
\end{table}
\normalsize

\section{DISCUSSION}
We present a technique for tracking the dynamical history of halos in an N-body simulation. Since many of these 
interactions are close flybys and not mergers,  we call this ensemble the halo interaction network. 
We find that most of the anomalies common to merger tree construction arise from limitations in the halo finding 
algorithm, and we correct for those in creating the halo interaction network.  For example, density contrast based methods
like \SUB are unable to track a subhalo when it reaches a denser central region. In most cases, such close pericenter passages
manifest as severe mass loss in the subhalo (with a fraction of the mass regained when the subhalo is farther away from the host center), 
and at times, the entire subhalo falls below the minimum particle limit
and is not identified at all. There are also issues of nomenclature -- in a near equal-mass major merger, it 
is unclear which halo is the dominant one, i.e., which halo is to be labeled the `main' halo. 
The algorithm we have presented here tracks several pathological cases
and corrects them. We find that even after fixing all such cases, there still remain halos that are parentless
or childless at a few-10\% level for halos comprised of $\lesssim$ 80 particles. Even with very finely spaced snapshot outputs, we can not 
eliminate all of these ill-behaved halos. Based on our findings, we recommend that semi-analytic studies based on N-body mergertrees use only those 
halos with at least $3-4$ times the formal particle limit. In our simulations, we formally require 20 particles to identify 
a halo, but all the results that have been reported here involve halos with at least 100 particles. 

For robust halos, our mergertree technique resolves anomalies related to halo definition and problems with halo identification; however,
the code can still be improved using maximal bipartite graph matching~\citep[e.g., ][]{HR73}. Given two sets of halos at
two different snapshots and an associated cost (say, the inverse of the binding energy rank) for matching any two of them, 
the maximal matching algorithms can be used to determine a match. Such a technique would produce the least number of 
anomalies and spurious initial assignments in a mergertree, and we will implement it in a future version of the mergertree code.
 
One major thrust of this paper is to include flyby interactions in our halo interaction network. In this paper, we present the first census of halo flybys across a wide 
range of halo mass and redshift. We classify a 
flyby using spatial {\em and} dynamical information. Conceptually, a grazing flyby occurs when a halo 
undergoes the transition from a main halo $\boldsymbol\rightarrow$subhalo$\boldsymbol\rightarrow$ a main halo and is fundamentally different
from mergers. In a simulation, it is relatively straightforward to identify such chains since the current kinematics 
and future behavior of a given interaction is fully determined.  We note that because flybys imply a population of galaxies that were once well within
the virial radius of the main halo but are now outside it, they may be related to so called `backsplash' galaxies. 
The morphological and kinematical features of these galaxies are likely to be 
different compared to other galaxies that are infalling for the first time~\citep{GKG05,MDS04,KLKMYGH11}. 
In some cases these flybys can even result in subhalos  switching their main halo and appear as `renegade' subhalos~\citep{KLDYGH11}. 
While there has been a significant amount of work related to `backsplash' galaxies in 
clusters~\citep[e.g.,][]{BNM00, SSSFM02, GKG05,MDS04,MSSS04, KLKMYGH11,MMR11}, unfortunately there are no systematic studies 
for lower mass halos. 

We suspect flyby interactions have been unappreciated in both observational and theoretical studies that attribute morphological/structural 
changes to mergers. For example, the pre-dominance of low-mass bulge-dominated galaxies ($M_{\rm stellar} \sim 10^{10}\,\Msun$) may be potentially 
explained by perturbations caused by flybys, since low-mass galaxies do not suffer enough mergers to 
explain the bulges~\citep{WJKBK09}. Indeed, flybys can cause a fractional change in the binding energy of the primary halo, $\dphi > 1\%$  
-- this may excite bars~\citep[see][and references therein]{BAHF04} and drive gas into the
central regions of the galaxy, creating a pseudo-bulge. Observationally, barred and unbarred galaxies can {\em not} be separated on some
inherent galaxy property (other than the actual presence/absence of a bar); this implies that the formation or destruction of a bar is likely
linked with some external process, such as a flyby~\citep{Se10}. In addition to forming bars/pseudo-bulges, flybys could 
create density enhancements via shocks in the hot halo gas of the primary galaxy. Such density enhancements will increase the 
cooling rates in the halo by $\sim$ an order of magnitude (assuming \brems and a factor of 4 increase in density). In principle this
can cause the hot gas to cool and replenish the disk~\citep{SH09}. In a similar spirit, flyby-induced star formation has been recently invoked to explain
the steepness of the [{$\alpha$}/Fe]-{$\sigma$} relation~\citep{CM11}. Flybys can also cause various other morphological features or transformations, e.g.,
spiral to S0 galaxy in groups~\citep{BC11}, spin flips in the inner halo~\citep{BF12} or spiral arms in the galactic disk~\citep[e.g.,][]{TF06}. 

\section{CONCLUSIONS}
In this paper we report on a new class of interactions -- flybys, that occur frequently. Most of these flybys are one-off events -- one halo delves 
within the virial radius of another main halo, separates at a later time and does not return. 
From our testbed simulation, we can see a hint that flybys dominate over mergers at very high redshift ($z\sim 14$) and have relatively shallower
dependence on $z$ compared to mergers (see Table ~\ref{table:fits}). We find that most flybys are one-off events and about 70\% of the 
flybys do not ever return. Flybys are then, a largely-ignored type of interaction that can potentially transform galaxies. Unfortunately, 
most semi-analytic methods of galaxy formation are designed only to use mergers and thus can not account for the
effects of flybys directly. In principle, a flyby can create long-lasting features to appear in the secondary halo and affect the 
evolution of that halo. In general, slow flybys cause a larger perturbation compared to a fast one, and such features can persist even when 
the perturbing halo has moved far away~\citep{VW00}. If flybys cause perceptible changes in galactic structure,
then those deviations should manifest in the oldest galaxies. Upcoming space telescopes like {\it JWST}
will target high redshift ($z\sim7$) galaxies to unravel clues for galaxy formation and may be able to detect flyby signatures.

In conducting our flyby census, our testbed simulation did not have adequate resolution to capture the perturbative effect of a flyby on the dynamics of the halo central 
region. We are currently exploring the effect of flyby interactions on galaxy structure and morphology using both high resolution N-body simulations 
and semi-analytic estimates (Sinha \& Holley-Bockelmann, in prep).  

\acknowledgments
This work was conducted in part using the resources of the Advanced Computing Center for Research and Education at Vanderbilt 
University, Nashville, TN.  We also acknowledge support from the NSF Career award AST-0847696.  Some of the simulations were run on the 
Pleiades cluster at NASA Advanced Supercomputing Division. We would like to thank the anonymous referee for helpful comments.
MS would like to thank Frazer Pearce and the organizers of the workshop  `Haloes Going MAD' for motivating some sections in 
this paper. MS would also like to thank H.R.~Madhusudan for pointing out bipartite graph matching algorithms.

\bibliographystyle{apj}
\bibliography{Biblio-Database}

\begin{appendix}

\section{HALO GENEALOGY: FINDING PROPER PARENT-CHILDREN PAIRS}\label{sec:parents}
Our first step is to track the main halos through time. Since even simple isolated halos change mass and position in each snapshot, this is not a trivial task. 
Overall, we link main halos and subhalos with different techniques. In most cases, we can assign the proper parent for a main halo in the following
way:
\begin{itemize}
\item Gather all the particles in all main halos at two consecutive snapshots. Note that this includes the particles in the 
subhalos because some of these particles may have joined (or left) the main halo between these two redshifts.

\item Working backwards from a later snapshot, we compare the particle ids in one halo with the ids in all primary halos at the 
previous snapshot. For computational efficiency, once a particle id is matched across two snapshots, we store the actual subhalos at the 
particle level.

\item Assign the parent\footnotemark halo as the one with the largest number of common particles.

\item Repeat for all remaining primary halos at the later snapshot

\item If several child halos claim the same parent (i.e. the parent halo may have split apart), the parent is assigned to the most massive child halo. 
In other words, a parent may have at most one child halo in our scheme. In addition, one child halo can have at most one parent at this stage -- though
this may change at the subhalo matching step.
\end{itemize}

\footnotetext{Here a parent halo lies in the past. Note that Extended Press-Schechter usually has the opposite convention.}

Using this procedure, we assign $\sim 90\%$ of the main halos to a parent. 
If the halo is still unassigned, we treat it as though it were a subhalo. Subhalos are matched using a binding energy technique described below.

In the \SUB algorithm, a subhalo is a self-bound group of particles. Therefore, an acceptable parent-child pair
should have most of the highly bound particles in common.  We track the subhalo core~\citep[similar to][]{OH00,SW05,LB07,BS09} 
using a binding energy rank similar to \citet{BS09}. For every particle that is common between two halos, we compute the quality -- $\mathcal{Q}$, as:
\begin{equation}\label{eqn:berank}
\mathcal{Q} = \displaystyle \sum_{i \leq 100} \mathcal{R}_{i, {\rm prev}}^{-2/3} + \displaystyle \sum_{j \leq 100}\, \mathcal{R}_{j, {\rm next}}^{-2/3} \quad , 
\end{equation}
where \prev and \next refer to the halo at the previous and later snapshot and $\mathcal{R}$ is the binding energy rank  (1 for the most bound, 
2 for the next and so on) of the common particle; this choice weights the core of a halo more strongly than the outer regions. 
We limit the number of particles in $\mathcal{Q}$ to \Nmost, because in cases of severe tidal stripping, only the core remains -- the rest of the subhalo is 
spread to the main halo background. If no child subhalo is found, we search two later snapshots using this binding energy rank method. 
This yields a parent-child match for over 99\% of the halos. 

After the parent-child matching is done for all halos and subhalos, we manually check the assignments for validity and 
make adjustments if necessary. We found that the parent-child pair mis-assignments typically fell into two 
classes -- those caused by major mergers, and those caused by numerical resolution.
Major mergers caused $< 1\%$ of the mis-assignments while particle resolution resulted in the remainder of the poor parent-child pairs.
During a major merger, the process may  be so violent that a large parent subhalo is the only thing that survives relatively intact, 
while the main parent halo is mixed throughout the remnant. In this case, the parent halo 
and subhalo effectively 'swap' categories after the merger.  However, because our halo matching takes place before subhalo matching, 
our algorithm inaccurately assigns both the parent halo and the subhalo to the same main child halo after the merger. 
The consequence of this is that it eliminates an entire branch of the merger tree.  To correct the pairing, 
we search for this pattern in the parent-halo pairs, demote the main parent halo to a child subhalo, and reinstate the missing branch.

\subsection{Tracking the network of interactions}\label{sec:network}
The parent-finding step described in the previous section simply links one halo to another between two timesteps, 
where possible. To make the halo interaction network, we must both uniquely track every halo from its formation 
to its destruction (or to the present day if it survives) and categorize every dynamical interaction between two or 
more halos.  The procedure for tracking the halo is quite similar to our technique for pairing an individual parent 
and child halo, and amounts to constructing a merger tree; for completeness sake, however, we describe how we follow 
a halo through its lifespan. First,  for every child halo at a given snapshot, we identify all the parents that comprise 
the child -- these parent halos are all direct parents, but may each have joined the child halo at different redshifts. 
We choose the most massive parent as the primary halo, and assign unique halo IDs that are preserved as we step backwards 
in time along a particular halo branch. 

\begin{figure}
\centering
\includegraphics[scale=0.25,clip=true]{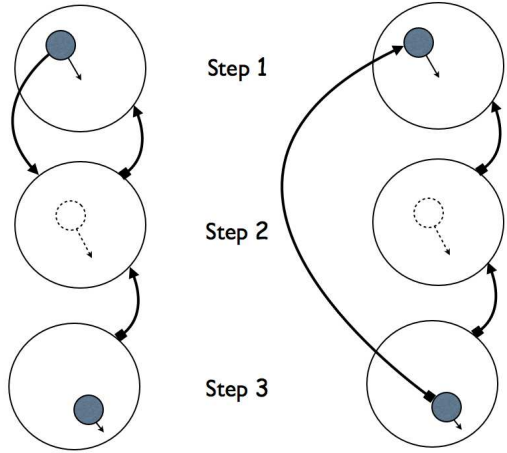}
\caption{\small Schematic for finding the parents of subhalos passing through
a dense halo center. Time increases vertically downwards. As the subhalo passes
through the central region, the density contrast becomes too low for \SUB to
identify it; consequently, original particles from the subhalo blend with
the halo background. At some later snapshot, when the subhalo is sufficiently far
from the center, \SUB identifies it as a subhalo (left). Left alone, the parent-matching algorithm
will lose the subhalo in step 1 to the main halo in step 2, and mistakenly identify a new parentless subhalo 
in step 3.  We remedy this by checking that the subhalo in step 3  can be found in the same halo in an 
earlier snapshot. }
\label{fig:passing_subs}
\end{figure}

While the halo-pairing step addressed most anomalies, there is one anomaly that can only be addressed during this step: when a 
subhalo passes through a dense region of the halo, it seems to disappear because \SUB cannot differentiate it from 
the background halo.  This means that when the subhalo reappears several snapshots later, it will seem brand new.  To 
graft these apparently new subhalo branches to their proper place in the network,  we first locate all the subhalos 
that do not have parents, and search among all subhalos up to five snapshots earlier for a suitable parent based on particle 
IDs. A potential match contains more than 50\% of the particles of the orphaned subhalo and is not a primary parent
of any other halo. Once the parent subhalo match is found, we attach the apparently orphaned subhalo branch to this parent subhalo (see
Fig~\ref{fig:passing_subs} for a schematic).

These orphans occur in a very small mass range - so fixing them does not dramatically increase the overall 99\% successful 
of parent-child assignments. The anomalies described in \citep{TDBCS09} are also present in our technique: 
\begin{itemize}
\item No child: None of the halo particles were found in any other halo, up to 3 snapshots into the future. 
\item No parent: Either none of the halo particles were found in another halo up to 5 snapshots in the past, or the potential parent
halo has already been matched with a child halo. 
\item Transients:  The halo appears to be a transient phenomenon - none of the halo particles are present in
any future or past halo. We determined this to be largely an effect of numerical noise, as we will discuss in the next section.
\end{itemize}

\subsection{Halo tracking and particle number}\label{section:resolution}
\begin{figure}
\centering
\includegraphics[width=0.45\textwidth,keepaspectratio=true,clip=true]{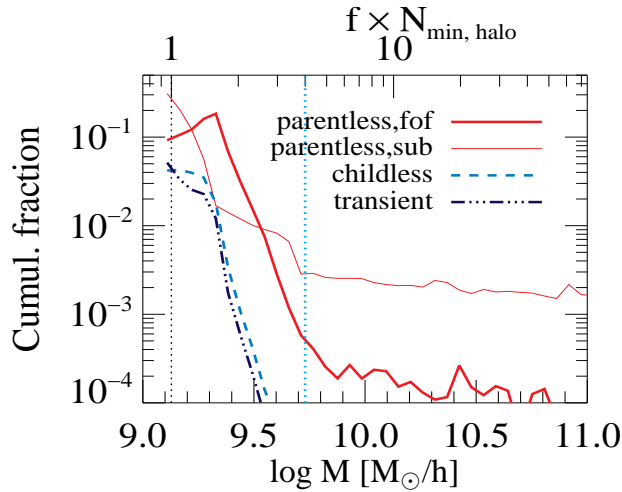}
\caption{\small The cumulative fraction of halo anomalies as a function of halo mass. 
The different lines represent parentless halo (thick solid), 
parentless subhalos (thin solid), childless (dashed) and transient (dash-dotted). 
The number of childless and transient halos drop off very sharply
with mass, suggesting that both are caused by numerical noise. The parentless halos
show a peak near our halo detection threshold -- these are all the newly-formed halos 
and hence do not have a parent. The parentless subhalos show a very gradual decline with mass, 
and can be interpreted as a combination of two effects: a) timing resolution -- a new halo forms and
is accreted by another halo to become a subhalo between two snapshots b) a  subhalo splits into two. 
The upper X-axis shows the number of resolution elements in a given halo mass; the cyan dotted line shows
our recommended minimum halo particle threshold to limit transient behavior at or below the 1\% level for reasonable 
timing resolution. The black dotted line shows the halo detection limit (20 particles), while the cyan line is the 4 
times that value. }
\label{fig:totn_res}
\end{figure}

In Fig.~\ref{fig:totn_res} we show the cumulative distribution, over all snapshots, for
the total number of parentless, childless and transient halos versus halo mass.
To generate the plot, we bin in mass for all the snapshots in a specified redshift range and 
count the total number of halos that are either parentless or childless. This is divided
by the total number of halos in the same range to give a fractional occurrence of the anomalies. 

It is clear that the number of childless and transient halos drop off very rapidly with particle number, and is essentially
zero for halos resolved with $\gtrsim 80$ particles, or about 4 times our nominal halo
detection threshold. This implies that the typical particle number threshold ($\sim 20$ particles) for resolving 
halos is much too small to produce robust results. A different halo finder and mergertree technique has found 
similar results (Dylan Tweed, private communication). With these results in mind, we find that a halo is unlikely to be noise 
once it is above $\sim 4$ times the canonical particle number threshold for halo detection. For the science results in the 
paper we will employ an even more conservative limit of 100 particles per halo, 5 times our halo detection threshold. 

\subsection{Halo tracking and time resolution}
Both parentless halos and parentless subhalos can occur; we expect most of the
parentless halos to be legitimate, newly formed halos. However, some of the 
parentless subhalos may actually be caused by a newly formed halo that is quickly accreted by another primary halo in between 
two snapshot outputs. To test this hypothesis, we increased the output resolution of the simulation
by an order of magnitude between $z=6-5$, resulting in a snapshot every few timesteps. 

\begin{figure}
\centering
\includegraphics[width=0.45\textwidth,keepaspectratio=true,clip=true]{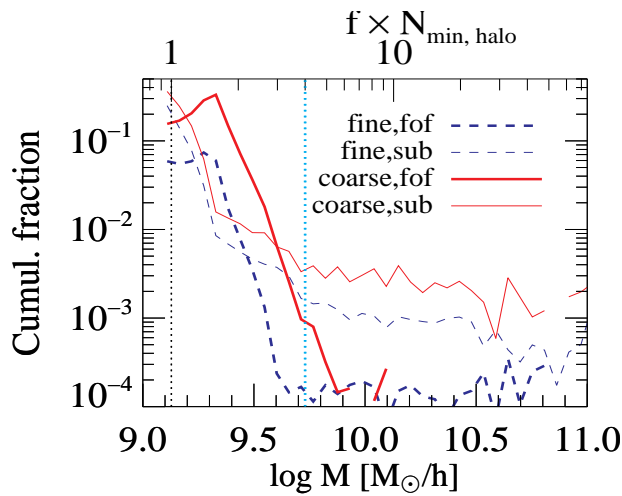}
\caption{\small This plot shows the dependence of parentless halos for the $z$ range $6-5$, as a function
of time resolution of snapshots. The red line shows our fiducial simulation while the blue shows the a re-simulation
with 10 times as many snapshots. The number of parentless halos in the fine simulation reduces
by factors of $4-5$ compared to the coarse one. While the number of parentless subhalos is also reduced in the fine
simulation, the drop is smaller. With such a high time resolution, we can reduce the fraction of parentless (sub)halos
to the few $\%$ level for halos that are only twice the detection limit. 
To achieve the fine time resolution shown here, a simulation would have to output about 2000 snapshots 
for $z=20$ to $0$ (assuming steps in $\log a$). The transients/childless halos show virtually no difference between
the two simulations and have been omitted from the plot. }
\label{fig:cumul_time_res}
\end{figure}

Fig.~\ref{fig:cumul_time_res} shows the percentage of halos that are parentless/childless in the redshift range $z=6-5$
for two simulations with differing snapshot output resolution. 
Our fiducial simulation, which we label as `coarse' in this plot, already has good timing resolution (see Section~\ref{sec:halo_identify}) 
but still produces anomalies at the $\sim 5-10\%$ level in the low mass halos. The `fine' simulation increases the snapshot output by a 
factor of 10 which reduces the level of anomalies by a factor of  $2-3$ over the entire mass range. 
Fig~\ref{fig:cumul_time_res} shows that the overall number of parentless subhalos dropped with this increased output frequency,  though 
never to the level of parentless halos. The slope for parentless subhalos are similar in the two simulations; in fact we note that the overall 
shape of the parentless subhalos mimics the cumulative mass function, albeit with a much lower amplitude. We expect that 
increasing the time resolution even further will reduce the amplitude, but a fraction of parentless subhalos will still remain. 
In the `fine' simulation,  the fraction of parentless subhalos reduces to the $\sim1\%$ at only twice the halo detection limit, but such accuracy
in the halo interaction network comes at the price of disk output -- a full-scale cosmological simulation would need $\sim 2000$ 
snapshots (assuming $\log a$ output spacing) to achieve the timing resolution of the `fine' simulation. The number of childless/transient halos 
is indistinguishable between the two simulations -- showing that they are indeed dominated by numerical noise. 

\subsection{Comparison of our technique to previous numerical work}
A mergertree simply identifies the links between components of a halo across redshifts. For particle-based codes, 
this means tracking halo particles by particle id across various snapshots. Nevertheless, the complexity of halo histories makes
the task non-trivial. As we noted earlier, most halo-finders use some form of a density contrast~\citep[e.g.,][]{SWTK01,KK09,KGKK99} 
to group particles into a subhalo (see VOBOZ~\citep{NGH05}, EnLink~\citep{SJ09}, HSF~\citep{MCSAB09} and Rockstar~\citep{BWW11} for different 
approaches). Such a technique fails to identify subhalos close
to the center of the host halo. Hence it is important to isolate such cases and re-assign the subhalos
that `disappear'. In this section, we compare the various strategies that have been employed to construct a mergertree based on
subhalos identified in N-body simulations. In particular, we compare our mergertree methods to those in
\citet{WBPKD02}, \citet{WCW09},  \citet{BS09} and \citet{TDBCS09}. 

In \citet{WBPKD02} a parent-child halo was assigned such that more than 50\% of the particles in the parent end up in 
the child along with those 50\% of particles constituting more than 70\% of the particles in the parent that end up in 
any halo. This technique is also labeled as the `most massive progenitor' method. Such an algorithm is prone to mis-assignment of 
parent-child halos in cases where a subhalo undergoes massive stripping or completely disappears inside a host halo. 

In \citet{WCW09}, the authors use the 20 most bound particles of a subhalo (that also occur in the potential child) 
to construct a parent-child pair. They compute the potential of these common particles based on all the particles
in the parent and weight against large increases in mass or scale-factor. However, the weighting function is linear in 
mass (Eqn. 2) while the potential itself is squared (Eqn. 1). If a subhalo appears to have dissolved into the
host halo, the large potential term could outweigh the limiting weighting function. As such, it is not obvious 
that this formulation would always pick the `right' subhalo as a child. In addition, since N-body simulations have discrete
outputs, weighting by the ratio of the scale-factors of the parent-child pair could change depending on the
output frequency of the same simulation. 

\citet{TDBCS09} specifically made the mergertree to account for the cases we have discussed in this paper. They assign parent-child
pairs by choosing the halo pairs with the maximum common mass as the main parent. In addition they identify (and remedy, if possible) three classes
of anomalies:  1) subhalos without a parent 2) subhalo that becomes the parent of a primary halo and 3) subhalo and host halo
swap identities. As we mentioned, we find two of these cases in our technique: anomaly 1 occurs under two circumstances -- when the time resolution of
the outputs is too coarse or when the subhalo appeared to have `dissolved' in the host halo at some earlier snapshot. We correct for the 
dissolved halos in our interaction network. Anomaly 2 does not appear
in our simulations since we first match main halos across two snapshots before matching subhalos. An analog of anomaly 3 occurs in our technique 
for major mergers -- when \SUB switches the main halo and dominant subhalo definition across two snapshots. We have outlined our method to fix
these cases in Section~\ref{sec:network}. Overall, the approach described here and those used by \citet{TDBCS09} agree qualitatively. 
We reiterate the cautionary statement from \citet{TDBCS09} that careful analysis has to be done to grow an `accurate' mergertree. Although the
anomalies that we have described and fixed represent $\lesssim 5\%$ of all halos, it is not clear to us what the cumulative effects of such 
anomalies are on semi-analytic approaches. 

Specifically the advantages in our mergertree technique are as follows:
\begin{description}
\item[First] It uses different strategies to locate descendants for main halos and subhalos. Descendants of main halos are assigned based on the 
highest number of common particles, subhalo descendants are assigned according to the binding energy rank (see Eqn.~\ref{eqn:berank}). The
rank computation is symmetric, i.e., a highly bound common particle always adds to the total rank. Using only common particles or binding 
energy rank leads to mis-assignments, e.g., a highly stripped subhalo will be lost even when the core survives or a main halo will get
assigned to a subhalo that happens to contains some of the most highly bound particles from the previous step. 
\item[Second] It carefully corrects the following pathological cases to the 0.1$\%$ level:
  \item[-] a subhalo and main halo switching definition across timesteps
  \item[-] a subhalo appearing to dissolve into the main halo due to massive tidal stripping (even
	though the core of the subhalo survives)
  \item[-] when a subhalo (including the core) seems to disappear for multiple snapshots while passing 
	close to the center of a main halo 
\end{description}

\end{appendix}

\end{document}